\documentclass[12pt]{article}
\usepackage{amssymb}
\usepackage{amsmath}
\usepackage{amscd}
\usepackage{latexsym}
\usepackage{dsfont}
\def\d{\delta}

\def\t{\theta}

\def\be{\begin{equation}}
\def\ee{\end{equation}}
\def\arr{\begin{array}{rll}}
\def\ea{\end{array}}
\def\bea{\begin{eqnarray}}
\def\eea{\end{eqnarray}}

\def\N2{$N{=}2$}

\def\&gt;{\rangle}
\def\&lt;{\langle}
\def\+{\dagger}
\def\={\ =\ }

\begin{document}
\renewcommand{\thefootnote}{\fnsymbol{footnote}}
\begin{titlepage}
\setcounter{page}{0}
\begin{flushright}
LMP-TPU--12/11  \\
\end{flushright}
\vskip 1cm
\begin{center}
{\LARGE\bf $\mathcal{N}=2$ supersymmetric extension of \\
\vskip 0.4cm
$l$-conformal Galilei algebra}\\
\vskip 2cm
$
\textrm{\Large Ivan Masterov\ }
$
\vskip 0.7cm
{\it
Laboratory of Mathematical Physics, Tomsk Polytechnic University, \\
634050 Tomsk, Lenin Ave. 30, Russian Federation} \\
{E-mail: masterov@tpu.ru}

\end{center}
\vskip 1cm
\begin{abstract} \noindent
$\mathcal{N}=2$ supersymmetric extension of the $l$-conformal Galilei algebra is
constructed. A relation between its representations in flat spacetime and in
Newton-Hooke spacetime is discussed. An infinite-dimensional generalization of the
superalgebra is given.
\end{abstract}


\vskip 0.5cm


\end{titlepage}

\renewcommand{\thefootnote}{\arabic{footnote}}
\setcounter{footnote}0

\noindent
{\bf 1. Introduction}\\
\noindent

The advance in understanding the non--relativistic version of the AdS/CFT
correspondence \cite{son,balasub} stimulates extensive investigation of
non--relativistic conformal algebras \cite{gal/2}-\cite{aiz} (for important earlier
studies see \cite{nied-sch}-\cite{Duval_3}). Algebras relevant for physical
applications in flat non--relativistic spacetime as well as in Newton-Hooke
spacetime (i.e. spacetime with universal cosmological repulsion/attraction
\cite{duvhor,gp}) belong to the family of the $l$-conformal Galilei algebras\footnote{In
modern literature
the algebra is also referred to as the conformal Galilei algebra with rational
dynamical exponent \cite{duvhor}, $N$-Galilean conformal algebra \cite{GomKam}, and
the spin-$l$ conformal Galilei algebra \cite{marttach}. In the present work we use
the terminology originally adopted in \cite{NdelO1,NdelO2}.} \cite{henkel97,NdelO1},
$l$ being a positive integer or half--integer.
Note that the way the temporal and spatial coordinates scale under dilatations
explicitly depends on $l$. Furthermore,
the number of vector generators grows with $l$ \cite{GM2,NdelO1}.

As is well known, the $l$-conformal extensions of the Galilei algebra and the
Newton-Hooke algebra are isomorphic \cite{NdelO1,nied1}.
If one makes a linear change of the basis in the conformal Newton-Hooke algebra $H
\to H \mp \frac{1}{R^2} K$, where
$H$ is the generator of time translations, $K$ is the generator of special conformal
transformations, and $R$ is the characteristic time, one arrives at the conformal
Galilei algebra. There are two subtle points regarding this isomorphism, however.
First, as far as dynamical realizations are concerned, the change of the bases is
actually a change of the Hamiltonian which alters the dynamics. Second,
the constant $R$ is dimensionful. If one has a model invariant under the conformal
Galilei group, in general, such a constant is not at our disposal. For this reason,
turning the conformal Galilei symmetry into the conformal Newton-Hooke symmetry may
happen to be problematic.
Yet, it is customary to speak about realizations of one and the same algebra in flat
spacetime and in the Newton-Hooke spacetime.

So far supersymmetric extensions of the $l$--conformal Galilei algebras and their
dynamical realizations have been studied in detail for $l=1/2$
(the Schr\"{o}dinger algebra) and $l=1$ (the conformal Galilei algebra). In
\cite{gaunlett} an $\mathcal{N}=1$ supersymmetric extension of the Schr\"{o}dinger
algebra was identified with the symmetry algebra of the non--relativistic
spin-$\frac{1}{2}$ particle. In \cite{llm} it was shown that the non--relativistic
limit of the Chern-Simons matter system in $(2+1)$ dimensions is invariant under
$\mathcal{N}=2$ Schr\"{o}dinger supersymmetry.
The systematic study of the Schr\"{o}dinger superalgebras was given in \cite{dh}.
Infinite dimensional Schr\"{o}dinger superalgebras was considered in paper \cite{henkel}. Many-body quantum mechanics invariant under $\mathcal{N}=2$ Schr\"{o}dinger
supersymmetry was studied in \cite{gm1,gal/1} (see also \cite{gal1,gal2}). Relations
between the Schr\"{o}dinger superalgebra and relativistic superconformal algebras were
discussed in \cite{yosida}--\cite{sorba}.
More recently, supersymmetric extensions of the conformal Galilei algebra were
extensively investigated by applying various non-relativistic contractions
\cite{Lukierski1}, \cite{sakaguchi}--\cite{Lukierski3}.

The purpose of this work is to construct an $\mathcal{N}=2$ supersymmetric extension
of the $l$-conformal Galilei algebra for the case of arbitrary $l$. We do this in
section 2. Representations of this superalgebra in flat spacetime and in
Newton-Hooke spacetime are considered in section 3. We also give a coordinate
transformation, which relates the representations.
An infinite-dimensional extension of the superalgebra is discussed in
section 4. We summarize our results and discuss possible further developments in
section 5.

Throughout the work summation over repeated indices is understood. Partial
derivatives with respect to the spatial coordinates $x_i$, the temporal coordinate
$t$ and the fermionic variables $\theta^{+}, \theta^{-}$ are denoted by
$\partial_i$, $\partial_t$, and $\partial_{\theta^{+}}, \partial_{\theta^{-}}$,
respectively. For fermions we use the left derivative.

\vspace{0.5cm}

\vspace{0.5cm}
\noindent
{\bf 2. $\mathcal{N}=2$ supersymmetric extension of the $l$-conformal Galilei
algebra}\\
\noindent

First let us recall the structure of the $l$-conformal Galilei algebra. It involves
the generator of time translations $H$, the generator of dilatations $D$, the
generator of special conformal transformations $K$, the generators of space
rotations $M_{ij}$, and a chain of vector generators $C_i^{(n)}$, $n=0,1,..,2l$. In
particular, for $n=0$ one obtains the generator of space translations, $n=1$ gives
the generator of Galilei boosts, while higher $n$ describe accelerations. The
non-vanishing structure relations read \cite{NdelO1}
\bea\label{algebra}
&&
[H,D]=H, \qquad\;\,  [H,C^{(n)}_i]=n C^{(n-1)}_i,\quad [D,C^{(n)}_i]=(n-l) C^{(n)}_i,
\nonumber\\[2pt]
&&
[H,K]=2 D, \qquad [D,K]=K, \qquad\quad\quad\; [K,C^{(n)}_i]=(n-2l) C^{(n+1)}_i,
\\[2pt]
&&
[M_{ij},C^{(n)}_k]=-\delta_{ik} C^{(n)}_j+\delta_{jk} C^{(n)}_i,\;
[M_{ij},M_{kl}]=-\delta_{ik} M_{jl}-\delta_{jl} M_{ik}+
\delta_{il} M_{jk}+\delta_{jk} M_{il}.
\nonumber
\eea
Note that $H$, $D$ and $K$ form the conformal algebra in one dimension $so(2,1)$.

In order to construct an $\mathcal{N}=2$ supersymmetric extension of this algebra,
we introduce a pair of supersymmetry generators $Q^+$ and $Q^-$, the superconformal
generators $S^+$ and $S^-$, fermionic partners of the vector generators $L_i^{(n)+}$
and $L_i^{(n)-}$ with $n=0,1,..,2l-1$, extra bosonic vector generators $P_i^{(n)}$
with $n=0,1,..,2l-2$, and the bosonic generator $J$ which corresponds to
$u(1)$--R--symmetry. It is assumed that the odd generators are antihermitian
conjugates of each other
\be
\left({Q^+}\right)^{\dagger}=-Q^-, \qquad \left({S^+}\right)^{\dagger}=-S^-, \qquad
\left(L^{(n)+}_i\right)^{\dagger}=-L_i^{(n)-}.
\ee
The bosonic operators $J$ and $P_i^{(n)}$ are taken to be antihermitian as well.

In addition to (\ref{algebra}) we impose the following structure relations
\bea\label{supergal}
&&
\{Q^+,Q^-\}=2iH,  \; \{Q^{\pm},S^{\mp}\}=2iD\pm J,
\;\;\{Q^{\pm},L_i^{(n)\mp}\}=iC_i^{(n)}\mp nP_i^{(n-1)},
\nonumber\\[2pt]
&&
\{S^+,S^-\}=2iK, \;\, [Q^{\pm},C_i^{(n)}]=nL_i^{(n-1)\pm},\;
\{S^{\pm},L_i^{(n)\mp}\}=iC_i^{(n+1)}\mp(n-2l+1)P_i^{(n)},
\nonumber\\[2pt]
&&
[H,S^{\pm}]=Q^{\pm},\quad\;\;\,  [Q^{\pm},P_i^{(n)}]=iL_i^{(n)\pm},
\quad\;\,[D,L_i^{(n)\pm}]=(n-l+1/2)L_i^{(n)\pm},
\nonumber\\[2pt]
&&
[K,Q^{\pm}]=-S^{\pm},\quad
[S^{\pm},C_i^{(n)}]=(n-2l)L_i^{(n)\pm},\;[D,P_i^{(n)}]=(n-l+1)P_i^{(n)},
\nonumber\\[2pt]
&&
[D,Q^{\pm}]=-\frac{1}{2}Q^{\pm},\;  [S^{\pm},P_i^{(n)}]=iL_i^{(n+1)\pm}, \;\;
[K,L_i^{(n)\pm}]=(n-2l+1)L_i^{(n+1)\pm},
\nonumber\\[2pt]
&&
[D,S^{\pm}]=\frac{1}{2}S^{\pm},\quad\; [H,L^{(n)\pm}_i]=nL_i^{(n-1)\pm},
\,\,[K,P_i^{(n)}]=(n-2l+2)P_i^{(n+1)}, \qquad\,\,\,
\nonumber\\[2pt]
&&
[J,Q^{\pm}]=\pm iQ^{\pm},  \quad [H,P^{(n)}_i]=nP_i^{(n-1)},
\quad\,[M_{ij},L^{(n)\pm}_k]=-\delta_{ik} L^{(n)\pm}_j+\delta_{jk} L^{(n)\pm}_i,
\nonumber\\[2pt]
&&
[J,S^{\pm}]=\pm iS^{\pm},\quad\, [J,L_i^{(n)\pm}]=\pm iL_i^{(n)\pm},
\;\;\;\,[M_{ij},P^{(n)}_k]=-\delta_{ik} P^{(n)}_j+\delta_{jk} P^{(n)}_i.
\eea
Note that $l=1/2$ reproduces the well known $\mathcal{N}=2$ Schr\"{o}dinger
superalgebra (see e.g. \cite{henkel,gm1} and reference therein).

\vspace{0.5cm}
\noindent
{\bf 3. Realizations in superspace}\\

First let us construct a realization of the superalgebra (\ref{supergal}) in flat
superspace. Introducing two Grassmann variables $\theta^+$ and $\theta^-$, which are
complex conjugates of each other ${\left(\theta^+\right)}^{\dagger}=\theta^-$, one
finds (see also a related work \cite{henkel})\footnote{Superconformal symmetries
parameterized by a discrete parameter were also considered in
\cite{pluschay}-\cite{pluschay2}.}
\begin{align}\label{gengal}
&
D=t\partial_t+lx_i\partial_i+\frac{1}{2}\theta^{-}\partial_{\theta^{-}}+\frac{1}{2}\theta^{+}\partial_{\theta^{+}},
&&
K=t^2\partial_t+2ltx_i\partial_i+t\theta^{-}\partial_{\theta^{-}}+t\theta^{+}\partial_{\theta^{+}},
\nonumber
\\[2pt]
&
S^{\pm}=t\theta^{\pm}\partial_t+it\partial_{\theta^{\mp}}+2l\theta^{\pm}x_i\partial_i+\theta^{\pm}\theta^{\mp}\partial_{\theta^{\mp}},
&& Q^{\pm}=i\partial_{\theta^{\mp}}+\theta^{\pm}\partial_t,
\nonumber
\\[2pt]
&
H=\partial_t, && J=i\theta^{+}\partial_{\theta^{+}}-i\theta^{-}\partial_{\theta^{-}},
\nonumber\\[2pt]
&
C_i^{(n)}=t^n\partial_i && n=0,..,2l,
\nonumber
\\[2pt]
&
P_i^{(n)}=\theta^{-}\theta^{+} t^n \partial_i && n=0,..,2l-2,
\nonumber\\[2pt]
&
L^{(n)\pm}_i=\theta^{\pm} t^n \partial_i, && n=0,..,2l-1,
\nonumber\\[2pt]
&
M_{ij}=x_i\partial_j-x_j\partial_i. && {}
\end{align}
Discarding the fermions one reproduces a realization of the $l$-conformal Galilei
algebra in \cite{NdelO1}.

In order to construct a realization of the superalgebra (\ref{supergal}) in
Newton-Hooke spacetime extended by fermionic variables,
we introduce an analogue of Niederer's transformation. Guided by the analysis in
\cite{GM2}, we first consider a coordinate transformation
\begin{align}\label{supernied}
&
t'=R\tan{(t/R)},&& t'=R\tanh{(t/R)},
\nonumber\\[2pt]
&
x'_i=(\cos{(t/R)})^{-2l} x_i,&& x'_i=(\cosh{(t/R)})^{-2l} x_i,
\nonumber\\[2pt]
&
\left(\theta^{\pm}\right)'=(\cos{(t/R)})^{-1}\theta^{\pm},&&
\left(\theta^{\pm}\right)'=(\cosh{(t/R)})^{-1}\theta^{\pm},
\end{align}
where the prime denotes coordinates parameterizing flat superspace. Here the
left/right column corresponds to Newton--Hooke spacetime with negative/positive
cosmological constant. Then we consider a linear change of the basis in the
$l$-conformal Galilei algebra
\be\label{isomorph}
H\rightarrow H\pm \frac{1}{R^2}K\mp\frac{1}{R}J, \qquad Q^{\pm}\rightarrow
Q^{\pm}\pm\frac{i}{R}S^{\pm},
\ee
where the upper/lower sign in the generator of time translations corresponds to
negative/positive cosmological constant. In the former case the two steps yield
\bea\label{genNH1}
&&
H=\partial_t-\frac{1}{R}\left(i\theta^{+}\partial_{\theta^{+}}-i\theta^{-}\partial_{\theta^{-}}\right),
\qquad
J=i\theta^{+}\partial_{\theta^{+}}-i\theta^{-}\partial_{\theta^{-}},
\nonumber\\[2pt]
&&
D=\frac{1}{2}R\sin(2t/R)\partial_t+l\cos(2t/R)x_i\partial_i+\frac{1}{2}\cos(2t/R)\theta^{-}\partial_{\theta^{-}}+\frac{1}{2}\cos(2t/R)\theta^{+}\partial_{\theta^{+}},
\nonumber\\[2pt]
&&
K=R^2(\sin(t/R))^2\partial_t+lR\sin(2t/R)x_i\partial_i+\frac{R}{2}\sin(2t/R)\theta^{-}\partial_{\theta^{-}}+\frac{R}{2}\sin(2t/R)\theta^{+}\partial_{\theta^{+}},
\nonumber
\eea
\bea
&&
Q^{\pm}=ie^{\frac{it}{R}}\partial_{\theta^{\mp}}+\theta^{\pm}e^{\frac{it}{R}}\partial_t+\frac{2il}{R}e^{\frac{it}{R}}\theta^{\pm}x_i\partial_i+
\frac{i}{R}e^{\frac{it}{R}}\theta^{\pm}\theta^{\mp}\partial_{\theta^{\mp}},
\nonumber\\[2pt]
&&
S^{\pm}=R\sin(t/R)\theta^{\pm}\partial_t+iR\sin(t/R)\partial_{\theta^{\mp}}+2l\cos(t/R)\theta^{\pm}x_i\partial_i+\cos(t/R)\theta^{\pm}\theta^{\mp}\partial_{\theta^{\mp}},
\nonumber\\[2pt]
&&
C_i^{(n)}=R^n (\sin(t/R))^n (\cos(t/R))^{2l-n}\partial_i, ~ \qquad \qquad\;\;
n=0,1,..,2l,
\nonumber\\[2pt]
&&
L_i^{(n)\pm}=\theta^{\pm}R^n (\sin(t/R))^n(\cos(t/R))^{2l-n-1}\partial_i, ~ \qquad\,
n=0,1,..,2l-1,
\nonumber\\[2pt]
&&
P_i^{(n)}=\theta^{-}\theta^{+}R^n (\sin(t/R))^n(\cos(t/R))^{2l-n-1}\partial_i,
\qquad  n=0,1,..,2l-2,
\nonumber
\\[2pt]
&&
M_{ij}=x_i\partial_j-x_j\partial_i.
\eea
In the latter case one finds
\bea\label{genNH2}
&&
H=\partial_t+\frac{1}{R}\left(i\theta^{+}\partial_{\theta^{+}}-i\theta^{-}\partial_{\theta^{-}}\right),
\nonumber\\[2pt]
&&
Q^{\pm}=i(\cosh(t/R)+i\sinh(t/R))\partial_{\theta^{\mp}}+\theta^{\pm}(\cosh(t/R)+i\sinh(t/R))\partial_t+
\nonumber\\[2pt]
&&
+\frac{2l}{R}(\sinh(t/R)+i\cosh(t/R))\theta^{\pm}x_i\partial_i+\frac{1}{R}(\sinh(t/R)+i\cosh(t/R))\theta^{\pm}\theta^{\mp}\partial_{\theta^{\mp}},
\eea
while other generators follow from those in (\ref{genNH1}) by changing the
trigonometric functions with the hyperbolic ones. In the flat space limit $R\rightarrow \infty$ the generators
(\ref{genNH1}), (\ref{genNH2}) reproduce (\ref{gengal}).

\vspace{0.3 cm}
\noindent
{\bf 4. Infinite-dimensional extension}\\
\noindent

The $l$-conformal Galilei algebra (\ref{algebra}) admits an infinite-dimensional
Virasoro--Kac--Moody--type extension \cite{marttach,GM2}. Let us extend the analysis
in \cite{GM2} to supersymmetric case.

Consider a set of operators
\bea\label{inf1}
&&
K^{(n)}=t^{n+1}\partial_t+l(n+1)t^{n}x_i\partial_i+\frac{1}{2}(n+1)t^n
\t^{+}\partial_{\t^{+}}+\frac{1}{2}(n+1)t^n \t^{-}\partial_{\t^{-}},
\nonumber\\[2pt]
&&
F^{(n)\pm}=it^{n+1}\partial_{\t^{\mp}}+\t^{\pm}t^{n+1}\partial_t+2l(n+1)t^n \t^{\pm}
x_i\partial_i+(n+1)t^{n}\t^{\pm}\t^{\mp}\partial_{\t^{\mp}},
\nonumber\\[2pt]
&&
C_i^{(n)}=t^{n}\partial_i,\qquad L_i^{(n)\pm}=\t^{\pm}t^n\partial_i,\qquad
P_i^{(n)}=\t^{-}\t^{+}t^n\partial_i,
\nonumber\\[2pt]
&&
J^{(n)}=it^n\t^{+}\partial_{\t^{+}}-it^n
\t^{-}\partial_{\t^{-}}-2lnt^{n-1}\t^{-}\t^{+}x_i\partial_i,\qquad
M_{ij}^{(n)}=t^n(x_i\partial_j-x_j\partial_i),
\eea
where $n$ is an arbitrary integer. It is straightforward to verify that
$K^{(-1)}$, $K^{(0)}$ and $K^{(1)}$ reproduce $H$, $D$ and $K$. $F^{(-1)+}$ and
$F^{(0)+}$ give $Q^+$ and $S^+$, while $F^{(-1)-}$ and $F^{(0)-}$ yield $Q^-$ and
$S^-$ which we displayed above in (\ref{gengal}). In order to close the algebra, one
has to further extend the set of generators (\ref{inf1}) to include
\bea\label{inf3}
&&
M_{ij}^{1(n)\pm}=\t^{\pm}t^n (x_i\partial_j-x_j\partial_i),\; \;
M_{ij}^{2(n)}=\t^{-}\t^{+}t^n (x_i\partial_j-x_j\partial_i).
\eea

The structure relations of the infinite-dimensional superalgebra read
\bea
&&
\{F^{(n)\pm},F^{(m)\mp}\}=2iK^{(n+m+1)}\pm(m-n)J^{(n+m+1)},[K^{(n)},K^{(m)}]=(m-n)K^{(n+m)},
\nonumber\\[2pt]
&&
[K^{(n)},F^{(m)\pm}]=(m-n/2+1/2)F^{(n+m)\pm},\qquad\quad\;\;[K^{(n)},J^{(m)}]=mJ^{(m+n)},
\nonumber\\[2pt]
&&
[K^{(n)},C^{(m)}_i]=(m-l(n+1))C^{(n+m)}_i,\qquad\qquad\quad\;\;[K^{(n)},M_{ij}^{(m)}]=m
M_{ij}^{(m+n)},
\nonumber
\eea
\bea\label{infext1}
&&
[K^{(n)},L^{(m)\pm}_i]=(m+(1/2-l)(n+1))L^{(n+m)\pm}_i,\quad\,[F^{(n)\pm},P_i^{(m)}]=\pm iL_i^{(n+m+1)\pm},
\nonumber\\[2pt]
&&
[K^{(n)},P^{(m)}_i]=(m+(1-l)(n+1))P^{(n+m)}_i,\qquad\quad [F^{(n)\pm},M_{ij}^{(m)}]=mM_{ij}^{1(m+n)\pm},
\nonumber\\[2pt]
&&
[F^{(n)\pm},C_i^{(m)}]=(m-2l(n+1))L^{(n+m)\pm}_i,\qquad\qquad\, [F^{(n)\pm},M^{2(m)}_{ij}]=\pm iM_{ij}^{1(n+m+1)\pm},
\nonumber\\[2pt]
&&
[K^{(n)},M_{ij}^{1(m)\pm}]=(m+n/2+1/2)M_{ij}^{1(n+m)\pm},\qquad\,[J^{(n)},F^{(m)\pm}]=\pm i F^{(n+m)\pm},
\nonumber\\[2pt]
&&
[K^{(n)},M_{ij}^{2(m)}]=(m+n+1)M_{ij}^{2(n+m)},\qquad\qquad\quad\;[J^{(n)},C^{(m)}_i]=2lnP^{(n+m-1)}_i;
\nonumber\\[2pt]
&&
[M_{ij}^{(n)},C_k^{(m)}]=\d_{jk} C_i^{(n+m)}-\d_{ik}
C_j^{(n+m)},\qquad\qquad\quad[J^{(n)},L_i^{(m)\pm}]=\pm
iL_{i}^{(n+m)\pm},
\nonumber\\[2pt]
&&
\{F^{(n)\pm},M_{ij}^{1(m)\mp}\}=i M_{ij}^{(n+m+1)}\mp(n+m+1)M_{ij}^{2(m+n)},[J^{(n)},M_{ij}^{1(m)\pm}]=\pm i M^{1(n+m)\pm}_{ij},
\nonumber\\[2pt]
&&
\{F^{(n)\pm},L_i^{(m)\mp}\}=iC_i^{(n+m+1)}\pm((2l-1)(n+1)-m)P_i^{(n+m)},
\nonumber\\[2pt]
&&
[M_{ij}^{1(n)\pm},C^{(m)}_k]=[M_{ij}^{(n)},L^{(m)\pm}_k]=\d_{jk}
L_i^{(n+m)\pm}-\d_{ik} L_j^{(n+m)\pm},
\nonumber\\[2pt]
&&
[M_{ij}^{2(n)},C^{(m)}_k]=\pm\{M^{1(n)\mp}_{ij},L^{(m)\pm}_k\}=[M_{ij}^{(n)},P^{(m)}_k]=\d_{jk}
P_i^{(n+m)}-\d_{ik} P_j^{(n+m)},
\nonumber\\[2pt]
&&
[M_{ij}^{(n)},M_{kl}^{(m)}]=-\delta_{ik} M_{jl}^{(n+m)}-\delta_{jl}
M_{ik}^{(n+m)}+\delta_{il} M_{jk}^{(n+m)}+\delta_{jk} M_{il}^{(n+m)},
\nonumber\\[2pt]
&&
[M_{ij}^{(n)},M_{kl}^{1(m)\pm}]=-\delta_{ik} M_{jl}^{1(n+m)\pm}-\delta_{jl}
M_{ik}^{1(n+m)\pm}+\delta_{il} M_{jk}^{1(n+m)\pm}+\delta_{jk} M_{il}^{1(n+m)\pm},
\nonumber\\[2pt]
&&
[M_{ij}^{(n)},M_{kl}^{2(m)}]=\{M_{ij}^{1(n)-},M_{kl}^{1(m)+}\}=\delta_{il}
M_{jk}^{2(n+m)}+\delta_{jk} M_{il}^{2(n+m)}-\delta_{ik} M_{jl}^{2(n+m)}-\delta_{jl}
M_{ik}^{2(n+m)}.
\nonumber\\[2pt]
&&
\eea

From these structure relations it follows that the generators $K^{(n)}$,
$F^{(n)\pm}$, $J^{(n)}$ form the $\mathcal{N}=2$ Neveu-Schwarz subalgebra
\cite{brink}.

Note that an infinite--dimensional Schr\"{o}dinger-Neveu-Schwarz superalgebra
$sns^{(N)}$  with $\mathcal{N}$ supercharges was considered in \cite{henkel}. The
superalgebra above can be viewed as a generalization of $sns^{(2)}$ to the case of
arbitrary dimension and arbitrary value of $l$.

\vspace{0.3cm}

\noindent
{\bf 5. Summary}\\
\noindent

To summarize, in this paper we have constructed an $\mathcal{N}=2$ supersymmetric
extension of the $l$-conformal Galilei algebra and its realizations in flat
spacetime and in Newton-Hooke spacetime. A coordinate transformation which links the
realizations was given.
An infinite--dimensional extension was proposed.

Let us discuss possible further developments of the present work. In
\cite{GomKam,AGM_DR} dynamical realizations of the $l$-conformal Galilei
algebra (\ref{algebra}) were considered. It would be interesting to extend the
analysis to the supersymmetric case.
As was shown in \cite{GM2}, the $l$-conformal Galilei algebra admits a central
extension for any $l$. It would be interesting to classify admissible central
extensions for the superalgebras proposed in this work.

\vspace{0.5cm}

\noindent{\bf Acknowledgements}\\

\noindent
We thank A. Galajinsky for helpful discussions and P. A. Horv\'{a}thy for a useful correspondence. This work was supported by the
Dynasty Foundation, RF Federal Program "Kadry" under contracts  16.740.11.0469,
P691, MSE Program "Nauka" under contract 1.604.2011, RFBR grant 12-02-00121 and LSS
grant 224.2012.2.

\end{document}